\documentclass[aps,prl,twocolumn,showpacs]{revtex4}
\usepackage{epsfig}
\usepackage{bm}

\begin{document}

\title{Quantum Spin Tomography in Ferromagnet-Normal Conductors}
\author{P. Samuelsson$^1$ and Arne Brataas$^2$}
\affiliation{$^1$Division of Mathematical Physics, Lund University, Box 118, S-221 00
Lund, Sweden}
\affiliation{$^2$Department of Physics, Norwegian University of Science and Technology,
NO-7491 Trondheim, Norway}

\begin{abstract}
We present a theory for a complete reconstruction of non-local spin
correlations in ferromagnet-normal conductors. This quantum spin tomography
is based on cross correlation measurements of electric currents into
ferromagnetic terminals with controllable magnetization directions. For normal
injectors, non-local spin correlations are universal and strong. The correlations are
suppressed by spin-flip scattering and, for ferromagnetic injectors, by
increasing injector polarization.
\end{abstract}



\maketitle

Spintronics utilizes the electron spin in electronics applications and is an
important subfield of condensed matter physics. It is possible to create metallic or semiconducting hybrid
ferromagnet-normal conductor systems smaller than the spin-flip length 
\cite{mesospin,spindot}, yet semiclassically large. Topics of current
interest such as spin injection, precession, and relaxation 
\cite{mesospin,spindot,Johnsil}, spin Hall effects \cite{Spinhall}, current
induced magnetization excitations \cite{RalphStiles}, the reciprocal 
magnetization dynamics induced spin-pumping \cite{Arnerev}, spin based
transistors \cite{Dattadas}, and ferromagnet-superconductor
heterostructures \cite{FS} focus on the average non-equilibrium spin
accumulation and dynamics.

The correlations between injected spins in ferromagnet-normal conductor
systems have received much less attention. In two-terminal junctions, 
current correlations have been investigated in few-level quantum
dots \cite{Bulka} as well as semiclassically large
systems \cite{Twoterm,Arne}. 
The prime targets have been noise due to spin-flip scattering and
the super or sub poissionian nature of the auto correlations.

In multiterminal junctions, current cross correlations allow
investigations of non-local spin transport properties. Of main
interest has been the sign of the cross correlations, studied in
quantum dots \cite{CotBel}, diffusive \cite{BelZar} and
superconducting \cite{Taddei} systems and chaotic cavities
\cite{SanLop}.  Moreover, in the context of entanglement of itinerant
spins, works on few-mode \cite{Rev1} and recently also semiclassical
\cite{dilorenzo,morten} conductors considered non-local detection
schemes with cross correlations between currents in non-collinear
ferromagnetic terminals. 

A fundamental and important question which has not been addressed is
if known non-local spin injection and detection schemes
\cite{Johnsil,mesospin} can be extended to identify non-local
spin-correlations. Imagine spins injected into a normal conductor and
detected at two different spatial locations by ferromagnetic
terminals.  What are the non-local spatial correlations between the
spins? Is it possible to completely characterize the correlations by
experimentally accessible electrical current correlations? We provide
answers to these questions for semiclassical systems: i) non-local
spin correlations are strong, and for normal injectors, universal and
ii) spin-correlations can be reconstructed by a sequence of
measurements of correlations of currents at ferromagnetic detectors
with controllable magnetization directions, a quantum spin tomography.

\begin{figure}[h]
\centerline{\psfig{figure=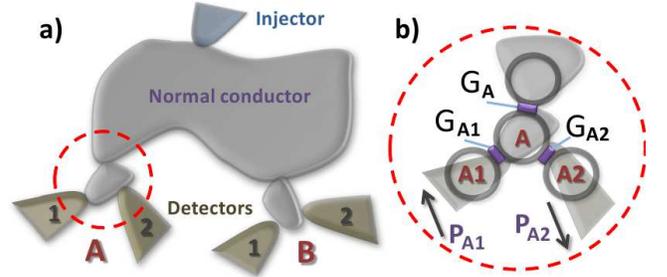,width=8.5cm}}
\caption{a) A normal conductor is connected to an injector biased at voltage 
$V$ and two detector nodes $A$ and $B$. The node $A$ ($B$) is coupled to
grounded ferromagnetic detector terminals $A1$ and $A2$ ($B1$ and $B2$). b)
Node $A$ is connected to the normal conductor, as well as nodes $A1$ and $A2$
via tunnel conductances $G_A$, $G_{A1}$, and $G_{A2}$, respectively. The
polarizations $\mathbf{P}_{A1}$ and $\mathbf{P}_{A2}$ of the contacts to the 
ferromagnetic terminals are in opposite directions.}
\label{fig1}
\end{figure}

We consider a semiclassically large, normal (metal or semi-) conductor
connected to a normal or ferromagnetic injector, biased at a voltage
$V$, and two spatially separated detectors, $A$ and $B$, see Fig
\ref{fig1}. Detector $A$ ($B$) consists of a normal node coupled to
grounded ferromagnetic terminals $A1$ and $A2$ ($B1$ and $B2$) via
tunnel contacts with conductances $G_{A1}$ and $G_{A2}$ ($G_{B1}$ and
$G_{B2}$). Throughout, conductances are dimensionless and in units of
the conductance quantum $2e^2/h$.  The detectors A and B probe
non-invasively the non-local spin correlations.

Let us first summarize and explain our main results i) and ii) for the
non-local correlated spin transport properties in the device in Fig.\
\ref{fig1}: First, combining scattering theory and a
Boltzmann-Langevin approach we derive an expression for the current
correlations $S_{AiBj}=(2e^{2}/h)\int_{0}^{eV}dEs_{AiBj}(E)$ with
\begin{equation}
s_{AiBj}=4G_{Ai}G_{Bj}\langle (\delta f_{A}^{c}+\mathbf{P}_{Ai}\cdot \delta 
\mathbf{f}_{A})(\delta f_{B}^{c}+\mathbf{P}_{Bj}\cdot \delta \mathbf{f}
_{B})\rangle_f,  
\label{Spin1}
\end{equation}
where $\mathbf{P}_{Ai}~(\mathbf{P}_{Bj})$ is the polarization of the
tunnel contact to terminal $Ai$ ($Bj$), $\delta \hat{f}_{A/B}=\delta
f_{A/B}^{c}\hat{1}+\delta \mathbf{f}_{A/B}\cdot
\hat{\boldsymbol{\sigma }}$ is the fluctuating part of the $2\times 2$
spin distribution matrix at $A/B$, $\hat{\boldsymbol{\sigma }}=
[\hat{\sigma}_{x},\hat{\sigma}_{y},\hat{\sigma}_{z}]$ is a vector of
Pauli matrices, and $\langle ..\rangle_f$ denotes the average over
fluctuations. The matrix $\delta \hat{f}_{AB}$, with elements $\delta
f_{AB}^{pq}=\langle \delta f_{A}^{p}\delta f_{B}^{q}\rangle_f$,
~$p,q\in \{c,x,y,z\}$, is the spin correlation matrix, describing the
irreducible, or exchange, correlations between spins at A and B.

We then show our result ii): $\delta \hat{f}_{AB}$ can be
reconstructed by a sequence of measurements of {\it e.g.} $S_{A1B1}$
with different settings of $\mathbf{P}_{A1}$ and
$\mathbf{P}_{B1}$. Importantly, this quantum spin tomography can be
performed for arbitrary (finite) magnitudes of the polarizations
$|\mathbf{P}_{A1}|$ and $|\mathbf{P}_{B1}|$ and spin-flip scattering
in the conductor.  Moreover, global spin symmetries limit the number
of finite elements of $\delta \hat{f}_{AB}$, allowing for a simplified
quantum spin tomography with fewer cross correlation measurements.

For a normal injector we derive a generic expression for $\delta
\hat{f}_{AB}$, with nonzero elements
\begin{equation}
\delta f_{AB}^{cc}=\delta f_{AB}^{0}/2,\hspace{0.25cm}\delta f_{AB}^{xx}=\delta
f_{AB}^{yy}=\delta f_{AB}^{zz}=\gamma \delta f_{AB}^{0}/2,  
\label{fABeq}
\end{equation}
where $\delta f_{AB}^{0}$ is the equal-spin correlator and
$\gamma $ quantifies the spin coherence in the conductor. $\gamma =1$
for a coherent system, $i.e.$ no spin-flip scattering, and $\gamma=0$
for a system with strong spin-flip relaxation. For a ferromagnetic
injector, the correlations depend on the properties of the conductor,
as shown below.

Inserting Eq. (\ref{fABeq}) into (\ref{Spin1}) gives a cross correlator
\begin{equation}
s_{AiBj}=2G_{Ai}G_{Bj}\delta f_{AB}^{0}[1+\gamma \mathbf{P}_{Ai}\cdot \mathbf{P}_{Bj}],
\label{crosscorr1}
\end{equation}
depending on the relative orientation of the polarizations
$\mathbf{P}_{Ai}$ and $\mathbf{P}_{Bj}$. This together with
Eq. (\ref{fABeq}) demonstrate our counter-intuitive result i):
\textit{any conductor with a normal injector displays strong and
universal non-local spin-correlations}. We note that for the current
cross correlator, similar results have been obtained in particular
geometries \cite{Rev1,dilorenzo,morten} with no spin-flip scattering,
$\gamma =1$.

We now describe the quantum spin tomography, starting for clarity with
the known properties \cite{Brataas} of the average spin distribution
matrix in node $A$, $\hat{f}_{A}=f_{A}^{c}\hat{1}+\mathbf{f}_{A}\cdot
\hat{\boldsymbol{\sigma }}$, where the real polarization vector
$\mathbf{f}_{A}=[f_{A}^{x},f_{A}^{y},f_{A}^{z}]$ with
$|\mathbf{f}_{A}|\leq 1$.  The average current is $I_{A1}=(e/h)
\int_{0}^{eV}dEi_{A1}(E)$ with \cite{Brataas}
\begin{equation}
i_{A1}=2G_{A1}\left[ f_{A}^{c}+\mathbf{P}_{A1}\cdot \mathbf{f}_{A}\right].
\label{curr}
\end{equation}
For the quantum spin tomography, we transform the orbital scheme
developed in Ref. \cite{tomo} to the spin degree of freedom and extend
it to account for arbitrary detector polarization. Formally, to
determine $f_{A}^{c}, \mathbf{f}_{A}$ four independent measurements of
the current are needed. The theoretically most convenient set
$\{I_{A1}^{(k)}\}$, $k=1-4$ has the polarizations
$\mathbf{P}_{A1}^{(1)}/P_{A1}=[0,0,1],\mathbf{P}
_{A1}^{(2)}/P_{A1}=[0,0,-1],\mathbf{P}_{A1}^{(3)}/P_{A1}=[1,0,0]$ and
$ \mathbf{P}_{A1}^{(4)}/P_{A1}=[0,1,0]$, where
$P_{A1}=|\mathbf{P}_{A1}|$, but other settings are also feasible. The
expression in Eq. (\ref{curr}) then allows writing
($\{k\}=\{c,x,y,z\}$)
\begin{equation}
f_{A}^{k}=\frac{\sum_{l=1}^{4}Q^{kl}_{A1}I_{A1}^{(j)}}{4G_{A1}e^{2}V/h},
\hspace{0.2cm}\frac{Q_{A1}}{P_{A1}}=\left( 
\begin{array}{cccc}
P_{A1}^{-1} & P_{A1}^{-1} & 0 & 0 \\ 
-1 & -1 & 2 & 0 \\ 
-1 & -1 & 0 & 2 \\ 
1 & -1 & 0 & 0
\end{array}
\right) 
\label{feq}
\end{equation}
Knowing the polarization $P_{A1}$ and the conductance $G_{A1}$ from
independent measurements, the spin-distribution matrix $\hat{f}_{A}$
is fully reconstructed by current measurements. Importantly, for a
normal injector, only $f_{A}^{c}$ is non-zero. For a ferromagnetic
injector, when the spin quantization axis along the direction of
polarization, only $f_{A}^{c}$ and $f_{A}^{z}$ are non-zero.

We then turn to the spin correlation matrix $\delta \hat f_{AB}$, with
the 16 real elements $\delta f_{AB}^{pq}$. This implies that we need
16 independent cross correlator measurements to determine all elements
$\delta f_{AB}^{pq}$ and reconstruct $\delta \hat{f}_{AB}$. From
Eq. (\ref{Spin1}) we obtain the formal relation between the
coefficients $\delta f_{AB}^{pq}$ and the cross correlators
\begin{equation}
\delta f_{AB}^{pq}=\frac{1}{8G_{A1}G_{B1}Ve^{3}/h}
\sum_{k,l=1}^{4}Q_{A1}^{pk}Q_{B1}^{ql}S_{A1B1}^{(k,l)},
\label{fABeq2}
\end{equation}
where $S_{A1B1}^{(k,l)}$ is the cross correlator with the detector terminal
setting $k$ at $A1$ and $l$ at $B1$. Here $Q_{B1}$ is obtained from $Q_{A1}$
by changing $P_{A1}$ to $P_{B1}$. 

For a normal injector, the requirement \cite{Been} of invariance of
$\delta \hat{f}_{AB}$ under any global spin rotation means that there
is only four non-zero elements $\delta f_{AB}^{cc}$ and $\delta
f_{AB}^{xx}=\delta f_{AB}^{yy}=\delta f_{AB}^{zz}$. For a
ferromagnetic injector (defining the spin quantization axis)
invariance of $\delta \hat{f}_{AB}$ under the global rotation
$\{|\!\!\uparrow \rangle ,|\!\!\downarrow \rangle \}\rightarrow
\{e^{i\phi }|\!\!\uparrow \rangle ,e^{-i\phi }|\!\!\downarrow \rangle
\}$ yields \cite{Wang} six non-zero elements $\delta
f_{AB}^{cc},\delta f_{AB}^{cz},\delta f_{AB}^{zc},\delta f_{AB}^{zz}$
and $\delta f_{AB}^{xx}=\delta f_{AB}^{yy}$.

From Eqs. (\ref{feq}) and (\ref{fABeq2}) the detector polarization
settings necessary to determine the non-zero components of $\hat f_A$
and $\delta \hat{f}_{AB}$ are found: For a normal injector, only
collinear polarizations at A and B are needed for both $\hat f_A$ and
$\delta \hat{f}_{AB}$.  For a ferromagnetic injector, for $\hat f_A$
the detector polarizations in addition have to be collinear with the
injector one. However, for $\delta \hat f_{AB}$ non-collinear
polarizations at A and B are necessary, {\it e.g.} both along the {\it x}
and {\it z} axis, since $\delta f_{AB}^{zz} \neq \delta
f_{AB}^{xx}=\delta f_{AB}^{yy}$. Importantly, for an unknown direction
of the injector polarization or two (or more) non-collinear
ferromagnetic injectors, the full tomographic scheme with detector
polarizations along all three axes {\it x,y} and {\it z} are required.

We will now detail our calculations, assumptions, and approximations. In
addition to the information given above, the normal conductor in Fig. \ref
{fig1} is connected to detector nodes $A$ and $B$ via tunnel barriers with
conductances $G_{A}$ and $G_{B}$. The two ferromagnetic
terminals $A1$ and $A2$ ~($B1$ and $B2$) have opposite directions of
polarization. We assume the limit of low temperature $kT\ll eV$. 
All conductances are much larger than unity.

It is assumed that the normal conductor consists of diffusive and/or
chaotic parts, allowing a semiclassical treatment of the orbital
properties. In contrast, spin is treated fully quantum mechanically.
Furthermore, scattering is elastic. Following the magnetoelectronic
circuit theory of Ref.  \cite{Brataas}, we discretize the system into
nodes connected via tunnel barriers, see Fig. \ref{fig1}. Each node
$\nu$, spatially much smaller than the spin-flip length, is
characterized by a $2\times 2$ distribution matrix with an average,
$\hat{f}_{\nu}$, and a fluctuating, $\delta \hat{f}_{\nu}$, part. To
ensure that the detectors do not influence the spin-properties of the
system, we require i) $G_{A}\ll G_{A1}+G_{A2}$ and $G_{B}\ll
G_{B1}+G_{B2}$ so that an electron entering \textit{e.g.} node $A$
from the conductor is emitted into $A1$ or $A2$ and do not return to
the conductor and ii) $G_{A1}\mathbf{P}_{A1}=-G_{A2}\mathbf{P}_{A2}$
and $G_{B1}\mathbf{P}_{B1}=-G_{B2}\mathbf{P}_{B2}$, which ensures that
no spin polarization is induced into the conductor from the
ferromagnetic terminals, \textit{i.e.} the measured spin signal arises
from the conductor exclusively and not from the detector circuits.

Deriving Eqs. (\ref{curr}) and (\ref{Spin1}), we first review \cite{Brataas}
the spin information present in the average spectral current $i_{A1}(E)$. In
the scattering approach \cite{Butrev}, with no particles incident from
terminal $A1$ in the bias window ($0\leq E\leq eV$), the spectral current is
\begin{equation}
i_{A1}=\sum_{n\sigma }\langle n_{A1,n}^{\sigma }\rangle ,\hspace{0.5cm}
n_{A1,n}^{\sigma }=b_{A1,n}^{\sigma \dagger }b_{A1,n}^{\sigma }
\label{curr2}
\end{equation}
where $b_{A1,n}^{\sigma \dagger }$ creates an electron on the
ferromagnetic side in the contact between $A1$ and $A$, in conduction
mode $n$ propagating into $A1$ and the energy-dependence is
suppressed. The spin quantization axis $\sigma =\uparrow ,\downarrow $
is along the direction of $\mathbf{P}_{A1}$. The creation operators
$b_{A1,n}^{\sigma \dagger }$ are related to the operators
$b_{Am}^{\tau \dagger }$ for electrons on the normal conductor side,
emitted from node $A$ towards $A1$, via the spin-dependent
transmission matrix of the normal-ferromagnetic interface $t_{A1}$
with elements $t_{A1,nm}^{\sigma \tau }$. Following Ref.
\cite{Brataas}, we make the semiclassical approximation that the spin
distribution matrix in node $A$ is independent on mode index,
\textit{i.e.}  $\langle b_{An}^{\sigma \dagger }b_{Am}^{\sigma
^{\prime }}\rangle = \hat{f}_{A}^{\sigma \sigma ^{\prime }}\delta
_{nm}$, giving
\begin{equation}
i_{A1}=\sum_{\sigma \tau }\hat{\mathcal{T}}_{A1}^{\tau \sigma }
\hat{f}_{A}^{\sigma \tau }=G_{A1}\mbox{tr}\left\{ (\hat{1}+\mathbf{P}_{A1}\cdot 
\hat{\boldsymbol{\sigma }})\hat{f}_{A}\right\} \,.  
\label{currmat}
\end{equation}
Here \cite{Brataas}
$\hat{\mathcal{T}}_{A1}=\sum_{nm}(\hat{t}_{A1,nm})^T(\hat{t}_{A1,nm})^*=G_{A1}(\hat{1}+\mathbf{P}_{A1}\cdot
\hat{\boldsymbol{\sigma }})$ where the elements of the $2\times 2$
matrix are $(\hat{t}_{A1,nm})_{\sigma \tau }=t_{A1,nm}^{\sigma \tau
}$. Eq. (\ref{currmat}) directly gives Eq. (\ref{curr}). Similar
relations hold for the average currents into $A2$, $B1$ and $B2$.

We then turn to the low frequency correlations between electrical
currents in \textit{e.g.}\ terminals $A1$ and $B1$, $S_{A1B1}=\int
dt\langle \Delta I_{A1}(0)\Delta I_{B1}(t)\rangle $. Scattering theory
\cite{Butrev} gives
\begin{equation}
s_{A1B1}=\sum_{nm,\sigma \tau }\left[ \langle n_{A1,n}^{\sigma
}n_{B1,m}^{\tau }\rangle -\langle n_{A1,n}^{\sigma }\rangle \langle
n_{B1,m}^{\tau }\rangle \right]   
\label{noise}
\end{equation}
where $n_{B1,m}^{\tau }=b_{B1,m}^{\tau \dagger }b_{B1,m}^{\tau }$ and
$b_{B1,m}^{\tau \dagger }$ creates an outgoing electron on the
ferromagnetic side, in conduction mode $m$ in $B1$ with spin
quantization axis along the direction of the magnetization
$\mathbf{n}_{B}$. Disregarding terms of second order in
$G_{A}/(G_{A1}+G_{A2})$ or $G_{B}/(G_{B1}+G_{B2})$, the operators
$b_{A1,n}^{\sigma \dagger }$ and $b_{B1,m}^{\tau \dagger }$ are
expressed in terms of the operators $b_{Ak}^{\sigma ^{\prime
}\dagger}$ and $b_{Bl}^{\tau ^{\prime }\dagger }$ and the scattering
amplitudes of the respective normal-ferromagnetic interfaces. Making
the semiclassical approximation that the non-local irreducible
correlator $\langle b_{An}^{\sigma \dagger }b_{Am}^{\sigma ^{\prime
}}b_{Bk}^{\tau \dagger }b_{Bl}^{\tau ^{\prime }}\rangle -\langle
b_{An}^{\sigma \dagger }b_{Am}^{\sigma ^{\prime }}\rangle \langle
b_{Bk}^{\tau \dagger }b_{Bl}^{\tau ^{\prime }}\rangle \equiv \delta
\hat{f}_{AB}^{\sigma \sigma ^{\prime },\tau \tau ^{\prime }}\delta
_{nm}\delta _{kl}$, we arrive at
\begin{eqnarray}
&&s_{A1B1}=\sum_{\sigma \sigma ^{\prime }\tau \tau ^{\prime }}
\hat{\mathcal{T}}_{A1}^{\sigma ^{\prime }\sigma
}\hat{\mathcal{T}}_{B1}^{\tau ^{\prime }\tau }\hat{f}_{AB}^{\sigma
\sigma ^{\prime },\tau \tau ^{\prime }}=G_{A1}G_{B1} \nonumber \\
&\times &\mbox{tr}\left\{ \left[ \left( \hat{1}+\mathbf{P}_{A1}\cdot
\hat{\boldsymbol{\sigma }}\right) \otimes \left(
\hat{1}+\mathbf{P}_{B1}\cdot \hat{\boldsymbol{\sigma }}\right) \right]
\delta \hat{f}_{AB}\right\}
\label{noisemat}
\end{eqnarray}
where $\hat{\mathcal{T}}_{B1}$ is obtained by changing all indices $A$
to $B$ in $\hat{\mathcal{T}}_{A1}$ and $\otimes$ is the tensor
product. Here we work in the basis $\{|\sigma\tau\rangle
\}=[|\!\uparrow \uparrow \rangle ,|\!\uparrow \downarrow \rangle
,|\!\downarrow \uparrow \rangle ,|\!\downarrow \downarrow \rangle ]$,
{\it i.e.}  the matrix elements $(\delta \hat{f}_{AB})_{|\sigma\tau\rangle,|\sigma'\tau'\rangle}=\delta
\hat{f}_{AB}^{\sigma\sigma',\tau\tau'}$. As is shown below, the
$4\times 4$ spin correlation matrix $\delta \hat{f}_{AB}=\langle
\delta \hat{f}_{A}\otimes \delta \hat{f}_{B}\rangle_f$, provides a
semiclassical interpretation of $\delta \hat{f}_{AB}$. This means that
Eq. (\ref{noisemat}) directly gives Eq. (\ref{Spin1}).  Moreover, the
expressions for the other correlators $S_{AiBj}$ can similarly be
given in terms of $\delta \hat{f}_{AB}$. This shows that
\textit{$\delta \hat{f}_{AB}$ contains all information about non-local
spin-correlations that can be obtained from cross correlations}.

To further investigate the properties of $\hat f_A, \hat f_B$ and
$\delta \hat f_{AB}$ we now turn to the spin-dependent
Boltzmann-Langevin approach of Ref.  \cite{Arne}. The average part of
the distribution matrix $\hat f_{\nu}$ at node $\nu$ is determined
from the condition of conservation of matrix currents into the node,
$\sum_{\mu} \hat i_{\nu\mu}=0$, $i_{\nu\mu}=-i_{\mu\nu}$. The $2\times
2$ matrix current between a normal node $\nu$ and a ferromagnetic or
normal node $\mu$ is $\hat i_{\nu\mu}=(G_{\nu\mu}/2)\left\{(\hat
1+\mathbf{P}_{\mu}\cdot \hat{\boldsymbol{\sigma}}), (\hat f_{\nu}-
\hat f_{\mu}) \right\}$ with $\{.,.\}$ the anti-commutator,
$G_{\nu\mu}$ the tunnel conductance between the nodes and
$\mathbf{P}_{\mu}$ the polarization vector of node $\mu$
($\mathbf{P}_{\mu}=0$ for a normal node). The distribution matrices
for normal and ferromagnetic terminal nodes are $\hat 1$ for biased
terminals and $0$ for grounded. This allow us to calculate the
distribution matrices of all nodes.

For the fluctuating part of the distribution matrix, we first note
that the total fluctuations of the matrix current $\Delta \hat
i_{\nu\mu}$ flowing between two nodes $\nu$ and $\mu$ is a sum of the
bare fluctuations $\delta \hat i_{\nu\mu}$ and $\delta \hat
{\mathcal{I}}_{\nu\mu}$ due to the fluctuating distribution
matrices. For $\nu$ normal and $\mu$ normal or ferromagnetic $\delta
\hat {\mathcal{I}}_{\nu\mu}=(G_{\nu\mu}/2)\left\{(\hat
1+\mathbf{P}_{\mu}\cdot \hat{\boldsymbol{\sigma}}),(\delta \hat
f_{\nu} -\delta \hat f_{\mu}) \right\}$. The requirement of matrix
current fluctuation conservation $\sum_{\mu} \Delta \hat i_{\nu\mu}=0$
then gives $\delta \hat f_{\nu}$ in terms of $\delta \hat
i_{\nu\mu}$. The bare fluctuations $\delta \hat i_{\nu\mu}$ at
different contacts are uncorrelated while for $\nu,\mu$ normal
$\langle \delta \hat i_{\nu\mu} \otimes \delta \hat i_{\nu\mu}
\rangle_f=(G_{\nu\mu}/2)\left[\hat f_{\nu}\otimes(\hat 1-\hat
f_{\mu})+\hat f_{\mu}\otimes (\hat 1-\hat f_{\nu}) \right]\hat W +
h.c.$ where $h.c.$ denote hermitian conjugate and the permutation
matrix $\hat W$ has nonzero elements
$W_{11}=W_{23}=W_{32}=W_{44}=1$. For $\nu$ normal and $\mu$
ferromagnetic we have $\langle \delta \hat i_{\nu\mu} \otimes \delta
\hat i_{\nu\mu} \rangle_f =(G_{\nu\mu}/2)\{([\hat
1+\mathbf{P}_{\mu}\cdot \hat{\boldsymbol{\sigma}}]
f_{\mu})\otimes(\hat 1- \hat f_{\nu})+ ([\hat 1+\mathbf{P}_{\mu}\cdot
\hat{\boldsymbol{\sigma}}][1-f_{\mu}])\otimes \hat f_{\nu}\}\hat W +
h.c.$. Here we used that ferromagnetic ({\it i.e}. terminal)
distributions do not fluctuate. From these relations any electrical
current correlator $\langle \Delta i_{\nu} \Delta i_{\mu}\rangle_f$,
with $\Delta i_{\nu}=\mbox{tr}\lbrack \Delta \hat i_{\nu}]$, can be
obtained.

Spin flip scattering is taken into account on the level of the
relaxation time approximation. This amounts to coupling each node $n$
to a spin-flip node $\varphi \nu$ with a tunnel contact with
conductance $G_{\varphi \nu} \propto 1/\tau _{\varphi \nu}$, with
$\tau_{\varphi \nu}$ the spin-flip time of the node, and requiring
conservation of \textit{electrical} current and current fluctuations
into the spin-flip node. Here we give the universal results of the
calculation, {\it i.e.} we consider an arbitrary normal conductor with any
amount of (spatially dependent) spin-flip scattering, the details of
the calculations are given elsewhere. First, by comparing the obtained
expression for the spectral cross correlators $s_{AiBj}=\langle \Delta
i_{A1}\Delta i_{B1}\rangle_f$ with Eq. (\ref{noisemat}) we conclude
that $\delta \hat f_{AB}=\langle \delta \hat f_A \otimes \delta \hat
f_B\rangle_f$, discussed above. Second, for normal injectors, we find
the generic form
\begin{eqnarray}
\delta \hat f_{AB}=(\delta f_{AB}^0/2) \left[(1-\gamma)\hat 1\otimes \hat
1+2\gamma\hat W\right]  
\label{corrgam}
\end{eqnarray}
This is just the result in Eq. (\ref{fABeq}).

Further insight is obtained by calculating the properties of the
simplest possible conductor, a single node \cite{spindot}. For a
normal injector we find the distribution function at {\it e.g.} A as $\hat
f_A=G_A/(G_{A1}+G_{A2})f\hat 1$, with $f=G/(G+G_A+G_B)$ the
distribution function of the conductor node and $G$ the
injector-conductor node conductance. This is independent on spin-flip
scattering. For the spin-correlation matrix we get the result in
Eq. (\ref{corrgam}) with $\delta
f_{AB}^0=f^3G_AG_B/[G(G_{A1}+G_{A2})(G_{B1}+G_{B2})]$ and
$\gamma=[1+\tau/\tau_{\varphi}]^{-1}$ with $\tau/\tau_{\varphi}=
G_{\varphi}/(G_A+G_B+G)$ the ratio of spin-flip and dwell times in the
central node.

For a ferromagnetic injector with polarization $\mathbf{P}_I$ the spin
distribution matrix at {\it e.g.} A has two non-zero components $(\hat
f_A)^{\uparrow\uparrow}\equiv f_A^{\uparrow}$ and $(\hat
f_A)^{\downarrow\downarrow}\equiv f_A^{\downarrow}$ with
$f_A^{\uparrow,\downarrow}=f[(1\pm P_I)(1\mp
P_If)+\tau_{\varphi}/\tau]/[(1-P_I^2f^2)+\tau_{\varphi}/\tau]$ with
$P_I=|\mathbf{P}_I|$. For the spin-correlation matrix, the full
expression, including spin-flip scattering, becomes very lengthy and
we only present the result for $\gamma=1$. This is $\delta
f_{AB}^{cc}=\delta f_{AB}^{zz}=\delta f_{AB}^0(c_++c_-)/2$, $\delta
f_{AB}^{cz}=\delta f_{AB}^{zc}=\delta \delta f_{AB}^0(c_+ -c_-)/2$ and
$\delta f_{AB}^{xx}=\delta f_{AB}^{yy}=\delta f_{AB}^0 c_0/2$ with
$c_{\pm}=(1\pm P_I)^2/(1 \pm P_If)^3$ and
$c_0=(1-P_I^2)/[1-(P_If)^2]$. Inserting this into Eq. (\ref{noisemat})
we get the cross correlator
\begin{equation}
S_{A1B1}=2G_{A1}G_{A1}\delta f_{AB}^{0}(c_I+c_0[1+\mathbf{P}_{A1}\cdot\mathbf{P}_{B1}])
\label{crosscorr}
\end{equation}
with $%
c_I=(1+P_{A1}^zP_{B1}^z)(c_++c_--2c_0)/2+(P_{A1}^z+P_{B1}^z)(c_+-c_-)/2$.
This clearly demonstrates that while a ferromagnetic injector leads to a
polarization of the conductor, it suppresses the spin-correlations.

In conclusion we have presented a scheme for quantum state tomography
of non-local spin correlations in normal-ferromagnetic
conductors. Non-local correlations are generically strong but
suppressed by spin-flip scattering and ferromagnetic injectors.

We acknowledge discussions with Daniel Huertas Hernando. This work was
supported by the Swedish VR and the Research Council of Norway, Grant
No 162742/V00.

\end{document}